\documentclass[aps,showpacs,amssymb]{revtex4}

\usepackage{epsf,graphicx}
\usepackage{supercite}
\usepackage[intlimits]{amsmath}
\usepackage{amsfonts}
\usepackage{psfrag}
\usepackage{dsfont}
\usepackage[dvips]{rotating}

\allowdisplaybreaks[1]
\newcommand {\hlf}{\frac{1}{2}}

\begin{document}

\title{Visibility of Cold Atomic Gases in Optical Lattices for Finite Temperatures}
\author{Alexander Hoffmann$^1$ and Axel Pelster$^2$}
\affiliation{$^1$Arnold Sommerfeld Center, Ludwig Maximilian Universit{\"a}t, 
Theresienstra{\ss}e 37, 80333 M{\"u}nchen, Germany\\
$^2$Fachbereich Physik, Universit{\"a}t Duisburg-Essen, 
Lotharstra{\ss}e 1, 47048 Duisburg, Germany}
\begin{abstract}
In nearly all experiments with ultracold atoms time-of-flight pictures are the only data available. In this 
paper we present an analytical strong-coupling calculation for those time-of-flight pictures  of bosons in an 
optical lattice in the Mott phase. This allows us to determine the visibility, which quantifies the contrast of 
peaks in the time-of-flight pictures, and we suggest how to use it as a thermometer.
\end{abstract}
\maketitle

\section{Introduction}

Systems of ultracold bosonic gases in optical lattices have recently become a popular research 
topic \cite{Lewenstein:AdP07,Bloch:RMP08}. After their theoretical suggestion \cite{Fisher:PRB89,Jaksch:PRL98} 
and first experimental realization \cite{Greiner:Nat02} it soon became clear that they represent model systems for 
solid-state physics with a yet unprecedented level of control. Both the periodic one-particle potential, which is 
superimposed with an additional harmonic trap for the purpose of confinement, and the two-particle interaction 
strength can be experimentally tuned with high precision \cite{Bloch:NaP05}. With this it is now possible to 
achieve strong correlations in these systems even though their particle densities are more than five orders of 
magnitude less than for air. In addition, due to the absence of impurities, they are viewed as idealized condensed 
matter systems which allow for a clear theoretical analysis \cite{Zwerger:ASS04,Zoller:AP(NY)05}.\\
Current research in optical lattices is driven by a number of cornerstone experiments which investigate, among other 
things, lower dimensionality \cite{Stoeferle:PRL04}, the formation of a distinct shell structure in the particle 
density \cite{Folling:PRL06}, and a more quantitative analysis of the time-of-flight interference patterns via their 
visibility \cite{Gerbier:PRA05}. Furthermore, dynamical aspects are now under crucial investigation \cite{Morsch:RmP06} 
as, for instance, the collapse and revival of the interference patters in nonequilibrium 
situations \cite{Greiner:Nat02.A,Kollath:PRL07,Schutzhold:ARX08}. Furthermore, using either bosonic or 
fermionic atoms in optical lattices reveals a contrasting bunching or antibunching behavior which can be fully 
attributed to the different quantum statistics of each atomic species \cite{Rom:Nat06}. Mixtures of bosons and 
fermions are also currently studied \cite{Esslinger:PRL06,Ospelkaus:PRL06,Refael:PRB08}, and this has even led to 
the creation of heteronuclear molecules \cite{Ospelkaus:PRL06.B}. Thus, in the near future it will be possible to 
test the theoretical prediction of new phases in optical lattices arising from an additional dipole-dipole 
interaction \cite{Damski:PRL03}. Another emerging line of research is the investigation of the effect of controlled 
disorder which can be created by several methods \cite{Lewenstein:AdP07}. It turns out that an additional frozen 
random potential leads to a Bose-glass phase due to the localization of bosons in the randomly distributed 
minima \cite{Fisher:PRB89,Krutitsky:NJP06,Krutitsky:08,Hofstetter:08}. 
Finally, it has been proposed to use ultracold bosons in optical lattices 
to realize a quantum computer \cite{Garcia:TRS03,Kay:NJP06,Treutlein:FPh06}.\\
An optical lattice is created by pairs of counter-propagating laser beams in all three dimensions. Loading bosons with 
mass $M$ in this laser field, the Stark effect leads to an effective periodic one-particle 
potential $V_{\rm ext}(\mathbf{x})$ with lattice spacing $a= \lambda / 2$ and laser wave length $\lambda$. 
Thus, the energy scale of the system is set by the recoil energy $E_R = \pi^2 \hbar^2 / 2 M a^2$, which is used 
to introduce dimensionless energies $\tilde{E}=E/E_R$. Neglecting the overall harmonic trapping potential, which 
is superimposed in order to spatially confine the system, spin-polarized bosons in an optical lattice can be described 
within the grand-canonical ensemble by the Bose-Hubbard Hamiltonian \cite{Fisher:PRB89,Jaksch:PRL98}
\begin{align}
\hat{H}=-J \sum_{<i,j>} \hat{a}_i^\dag \hat{a}_j + \sum_i \left( \frac{U}{2} \hat{a }_i^\dag \hat{a }_i^\dag 
\hat{a}_i \hat{a}_i -\mu \hat{a}_i^\dag \hat{a}_i \right) ,\label{eq:bhm} 
\end{align}
where $\hat{a }_i^\dag$ and $\hat{a }_i$ denote the standard creation and annihilation operators at site 
$i$, $\mu$ denotes 
the chemical potential, and the sum over $<i,j>$ includes only pairs of nearest neighbors. Both the hopping 
matrix element $J$ and the on-site interaction $U$ define the major energy scales of the system, which turn out to 
depend crucially on the strength $V_0$ of the laser field (see Appendix \ref{chp:para}). If $V_0$ is varied, the bosons 
can undergo a quantum phase transition. For small $V_0$ the hopping matrix element $J$ is large and the bosons can 
tunnel from site to site to explore the whole lattice. This leads to a superfluid state which is characterized by 
long-range correlations, a continuous excitation spectrum, and a finite compressibility. In the situation of large 
$V_0$ the hopping matrix element becomes negligibly small and the bosons can no longer tunnel to the neighboring sites, 
so the occupation number of the sites is fixed. This so-called Mott phase has no long-range correlation, shows a gap in 
the excitation spectrum, and is nearly
incompressible. In time-of-flight pictures these two phases can also be distinguished 
by their distinct interference patterns. In the superfluid phase the atoms are delocalized over the complete lattice. 
Therefore, according to the Heisenberg principle, their momentum uncertainty is small, which leads to sharp Bragg 
peaks in the time-of-flight pictures. In the Mott phase, on the other hand, all particles are strongly localized at 
lattice sites, leading to a large momentum uncertainty and ultimately to a uniform cloud during the expansion.\\
It is of particular interest to determine how the location of the transition from the superfluid to the Mott phase 
depends on the respective system parameters. Usually, one assumes that the temperature in the experiments is so low 
that thermal effects are completely negligible. In that case the phase boundary has been determined analytically within 
both a mean-field theory \cite{Fisher:PRB89,Oosten:PRA01} and a strong-coupling 
approach \cite{Freericks:PRB96,Damski:PRA06} as well as numerically by Monte-Carlo 
simulations \cite{Batrouni:PRL95,Capogrosso:PRB07,Capogrosso:PRA08}. Recently, with the field-theoretic concept 
of effective potential, the complete quantum phase diagram has been calculated essentially exact in excellent 
agreement with available numerical data \cite{Santos:ARX08}. Only  some theoretical work has been initiated to include 
thermal effects in a systematic way \cite{Krutitsky:NJP06,Buonsante:PRA04,Gerbier:PRL07,Cramer:ARX07}. However, until 
today, a reliable 
method of determining the temperature of the bosons in an optical lattice is not known \cite{Rom:Nat06}. 
Therefore, more experimental and theoretical studies are needed that aim at designing a thermometer for these systems.\\
This motivates the present paper where we investigate how the temperature affects the time-of-flight pictures for such 
a lattice system. We start in Section \ref{chp:corr} with determining perturbatively the correlation function for the 
homogeneous Bose-Hubbard model \eqref{eq:bhm} within the Mott phase. To this end we work out a hopping expansion up to 
second order at finite temperatures. Note that this hopping expansion is closely related to the random walk
expansion worked out in Refs.~\cite{Ziegler:94,Ziegler:02,Ziegler:03}.
The resulting correlation function is used in Section \ref{chp:tof} to 
qualitatively reconstruct time-of-flight absorption pictures which are taken after switching off the 
one-particle potential. From this we calculate in Section \ref{chp:visibility} the visibility, which quantifies the 
contrast of the time-of-flight pictures, determine how it changes with varying lattice depth $V_0$, and compare our 
results with experimental data. The Appendix summarizes more technical material. In Appendix \ref{chp:para} we 
determine how the Bose Hubbard parameters $J$ and $U$ depend on the lattice depth $V_0$. In particular, we show that 
the commonly used harmonic approximation \cite{Zwerger:JoO03,Albus:PRA03} deviates significantly from numerically 
determined results for large lattice depth. Finally, we work out in Appendix \ref{chp:shell} how an additional 
external harmonic confining potential fixes the average number of bosons per site. 
\section{Correlation function}\label{chp:corr}
The time-of-flight absorption pictures rely on the correlation function 
\begin{align}
\langle \hat{a }^\dag_i \hat{a }_j\rangle=\mathcal{Z}^{-1} {\rm Tr}\left\{\hat{a}^\dag_i \hat{a}_j 
e^{-\beta \Hat{H}} \right\} \label{eq:corrfunc}
\end{align}
with the partition function $\mathcal{Z}={\rm Tr}\left\{e^{-\beta \Hat{H}} \right\}$. As it is not possible to 
calculate this quantity analytically by exactly diagonalizing the Bose-Hubbard Hamiltonian, we have to employ a 
perturbative scheme. Therefore, we will restrict ourselves to the strong-coupling regime of the Mott phase, where the 
exactly solvable on-site part $\Hat{H}_0=\sum_i \left[ \frac{U}{2} \hat{n}_i ( \hat{n}_i-1) -\mu \hat{n}_i \right]$, 
with the number operator $\hat{n}_i=\hat{a}^\dag_i \hat{a}_i$, determines the unperturbed system and the hopping 
part $\hat{V}=-J \sum_{<i,j>} \hat{a}_i^\dag \hat{a}_j$ can be considered a small perturbation.
With this we can rewrite \eqref{eq:corrfunc} using the imaginary-time evolution operator in the Dirac 
picture $\hat{U}_{\rm D}(\tau,\tau') = e^{\hat{H}_0\tau/\hbar}e^{\hat{H}(\tau'-\tau)/\hbar} e^{-\hat{H}_0\tau'/\hbar}$ 
according to
\begin{align}
\langle \hat{a }^\dag_i \hat{a }_j\rangle=\mathcal{Z}^{-1} {\rm Tr}\left\{\hat{a}^\dag_i \hat{a}_j 
e^{-\beta \Hat{H_0}}\hat{U}_{\rm D}(\hbar\beta,0) \right\} \quad, \label{eq:corrU}
\end{align}
where also the partition function can be written in term of $\hat{U}_{\rm D}$:
\begin{align}
\mathcal{Z}={\rm Tr}\left\{ e^{-\beta \Hat{H_0}}\hat{U}_{\rm D}(\hbar\beta,0) \right\} \label{eq:ZU} \quad .
\end{align}
The Dirac time-evolution operator can now be expressed by the Dyson series:
\begin{align}
\hat{U}_{\rm D}(\tau,\tau')=1+\frac{-1}{\hbar}\int_{\tau'}^\tau d\tau_1 \hat{V}_{\rm D}(\tau_1)
+\left(\frac{-1}{\hbar}\right)^2 \int_{\tau'}^\tau d\tau_1 \int_{\tau'}^{\tau_1} d\tau_2 \hat{V}_{\rm D}(\tau_1)
\hat{V}_{\rm D}(\tau_2) +\ldots \quad . \label{eq:dyson}
\end{align}
With this we obtain from Eq.~(\ref{eq:ZU}) an expansion of the partition function in powers of the tunnel matrix 
element $J$. As the trace is only sensitive to the diagonal part and as one always needs in a cubic lattice an even 
number of steps to return to the starting point, all contributions with an odd power of $J$ must vanish leaving
\begin{align}
	\mathcal{Z}=\mathcal{Z}^{(0)}+J^2\mathcal{Z}^{(2)}+\ldots
\end{align}
with the coefficients
\begin{align}
\mathcal{Z}^{(0)}=&{\rm Tr} \left\{ e^{-\beta\hat{H}_0}\right\} \quad ,\label{eq:z0}\\
\mathcal{Z}^{(2)}=&\frac{1}{J^2\hbar^2}{\rm Tr} \left\{ e^{-\beta\hat{H}_0} \int_0^{\hbar\beta} d\tau_1 
\int_0^{\tau_1} d\tau_2 \hat{V}_{\rm D}(\tau_1)\hat{V}_{\rm D}(\tau_2)\right\} \label{eq:z1}\quad .
\end{align}
The correlation function \eqref{eq:corrU} is now calculated in the same manner by applying the Dyson 
series \eqref{eq:dyson}, so that we end up with a perturbation series in $J$:
\begin{align}
\langle \hat{a }^\dag_i \hat{a }_j\rangle= c_{ij}^{(0)}+J c_{ij}^{(1)}+J^2 c_{ij}^{(2)}+\ldots \quad.
\end{align}
Here the respective expansion coefficients read up to the second order in $J$:
\begin{align}
c_{ij}^{(0)}=&\frac{1}{\mathcal{Z}^{(0)}} {\rm Tr} \left\{ \hat{a}_i^\dag  \hat{a}_j 
e^{-\beta\hat{H}_0} \right\} \quad ,\label{eq:c0}\\
c_{ij}^{(1)}=&-\frac{1}{J \hbar \mathcal{Z}^{(0)}} {\rm Tr} \left\{ \hat{a}_i^\dag  \hat{a}_j 
e^{-\beta\hat{H}_0} \int^{\hbar\beta}_0 d\tau_1 \, \hat{V}_{\rm D}(\tau_1)\right\}\quad ,\label{eq:c1}\\
c_{ij}^{(2)}=&\frac{1}{J^2\hbar^2 \mathcal{Z}^{(0)}}{\rm Tr} \left\{ \hat{a}_i^\dag  \hat{a}_j  
e^{-\beta\hat{H}_0} \int_0^{\hbar\beta} d\tau_1 \int_0^{\tau_1} d\tau_2 \hat{V}_{\rm D}(\tau_1)
\hat{V}_{\rm D}(\tau_2)\right\} - \frac{ c_{ij}^{(0)} \mathcal{Z}^{(2)}} {\mathcal{Z}^{(0)}}\quad. \label{eq:c2}
\end{align}
Now we have to explicitly evaluate all the respective traces. To this end we use the occupation number 
basis $|\vec{n}\rangle=\prod_i |n_i\rangle$, in which the unperturbed system is diagonal, 
i.e. $\Hat{H}_0|\vec{n}\rangle=E_{\vec{n}}|\vec{n}\rangle$ with $E_{\vec{n}}=\sum_iE_{n_i}$, with the unperturbed 
single-site energies of the system given by $E_n=Un(n-1)/2-\mu n$.
Thus, representing the trace according to
\begin{align}
{\rm Tr} \{ \bullet\}=\sum_{\vec{n}}\langle \vec{n}|\bullet|\vec{n}\rangle \quad , \label{eq:trace}
\end{align}
the coefficients of the partition function (\ref{eq:z0}) and (\ref{eq:z1}) reduce to
\begin{align}
\mathcal{Z}^{(0)}=&\sum_{\vec{n}} e^{-\beta \sum_q E_{n_q}} \quad, \\
\mathcal{Z}^{(2)}=& \beta \sum_{<i,j>} \sum_{\vec{n}} \frac{n_i(n_j+1)} {E_{n_j+1}+E_{n_i-1}-E_{n_j}-E_{n_i}} 
e^{-\beta \sum_q E_{n_q} }\quad.
\end{align}
Analogous calculations yield the coefficients of the correlation function \eqref{eq:c0}--\eqref{eq:c2}. With the notation
\begin{align}
\langle \bullet \rangle_{k,l,\ldots}=\frac{ \sum_{n_k,n_l,\ldots} \bullet \; 
e^{-{\beta}\sum_{\nu=k,l,\ldots} E_{n_\nu}} }{\sum_{n_k,n_l,\ldots} e^{-{\beta}\sum_{\nu=k,l,\ldots} E_{n_\nu}}}
\end{align}
these can be written as
\begin{align}
c_{ij}^{(0)}=& \delta_{i,j} \langle n_i \rangle_i \label{eq:c0res}\quad ,\\
c_{ij}^{(1)}=& \frac{\delta_{d(i,j),1}}{U} \left\langle \frac{2 n_i(n_i+1)}{(n_i-n_j+1)(n_j-n_i+1)} 
\right\rangle_{i,j}\quad , \\
\begin{split}
c_{ij}^{(2)}=& \frac{\delta_{d(i,j),2}+2\delta_{d(i,j),\sqrt{2}}}{U^2} \left\langle 
\frac{n_i(n_j+1)(n_l+1)}{(n_j-n_i+1)(n_l-n_i+1)}  +\frac{n_in_j(n_l+1)}{(n_l-n_i+1)(n_l-n_j+1)}\right\rangle_{i,j,l}\\
&+\frac{z\beta\delta_{i,j}}{U} \left[ \left\langle \frac{n_i^2(n_l+1)}{n_l-n_i+1} \right\rangle_{i,l} 
- \langle n_i\rangle_i \left\langle  \frac{n_i(n_l+1)}{n_l-n_i+1} \right\rangle_{i,l} \right] \label{eq:c2res} \quad ,
\end{split}
\end{align}
where $d(i,j)$ is the distance in the Euclidean norm measured in units of the lattice spacing and $z=6$ is 
the number of next neighbors in a three-dimensional cubic lattice.
\section{Time-of-flight Absorption pictures}\label{chp:tof}
\begin{figure}[t]
\includegraphics{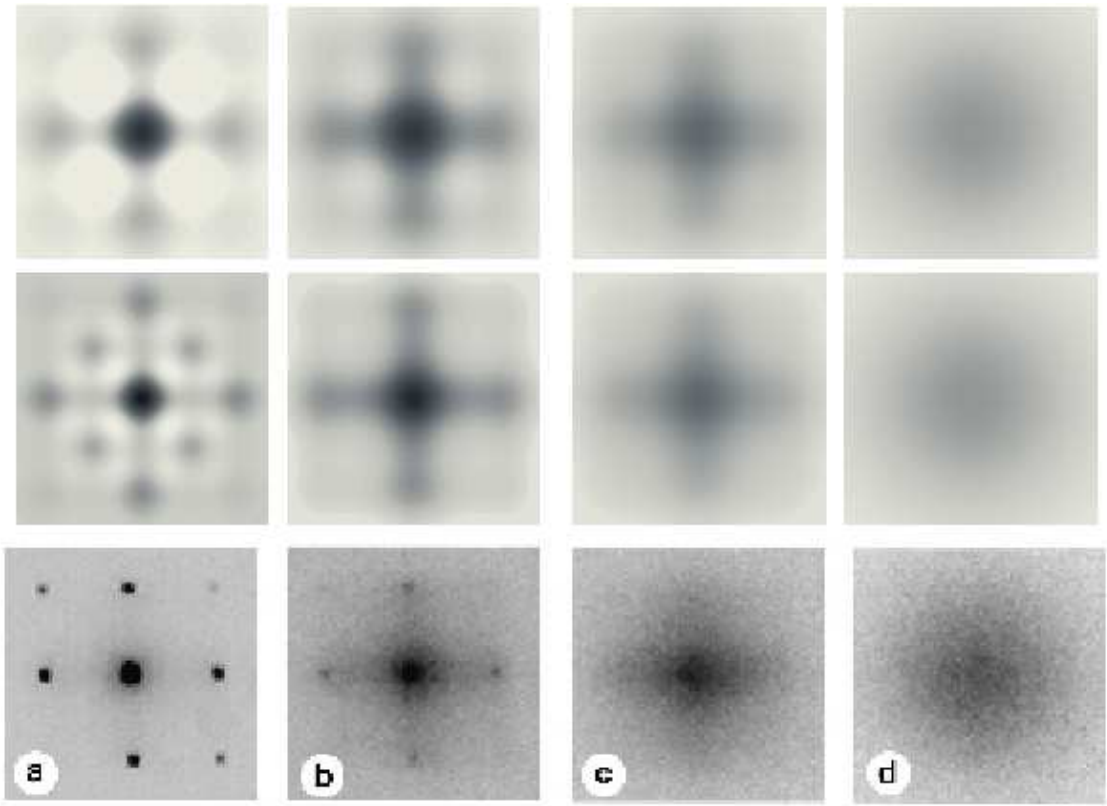}
\caption{Time-of-flight pictures for $\tilde{V}_0=8$ (column a), $\tilde{V}_0=14$ (column b), 
$\tilde{V}_0=18$ (column c), and $\tilde{V}_0=30$ (column d). First [second] row shows first-[second-]order 
calculation and the last row corresponds to the experimental data of Ref.~\cite{Gerbier:PRL05}.} \label{fig:tof}
\end{figure}

Measurements in most experiments on ultracold atomic gases are made as follows. The trapping potential is switched 
off allowing the gas to freely expand during a short time of flight $t$. Then an absorption picture is taken, which 
maps the particle density in real space to a plane. Due to the diluteness
we can neglect any interaction between the atoms, so the particles will approximately
move with a constant velocity given by their momentum at the moment of 
release from the trap. Furthermore, we will assume that the condensate in the trap is point-like, so that the 
distribution in real space is a mirror of the distribution in momentum space connected by the relation  
$\hbar{\bf k}=m{\bf r} /t$.
The momentum space distribution in the optical lattice is given by \cite{Karshunikov:PRA02}
\begin{align}
n({\bf k})=|w(\mathbf{k})|^2 S({\mathbf{k}})\quad ,
\end{align}
where the quasi-momentum distribution reads
\begin{align}
S({\mathbf{k}})=\sum_{i,j} e^{i {\bf k}({\bf r}_i-{\bf r}_j)} \langle\hat{a}^\dag_i \hat{a}_j\rangle
\end{align}
with the Fourier transform of the Wannier function being defined according to
\begin{align} 
w(\mathbf{k})=\int \frac{d^3x}{\sqrt{2\pi}^3}w(\mathbf{x}) e^{i\mathbf{kx}}. 
\end{align}
With \eqref{eq:corrU} and \eqref{eq:c0res}--\eqref{eq:c2res} we get the quasi-momentum distribution $S({\mathbf{k}})$ as
\begin{align}
S({\bf k},T)=S_0(T)+2\frac{J}{U} S_1(T) \sum_{i=1}^3 \cos(k_i a)+2\frac{J^2}{U^2} S_2(T) 
\left[-3+\sum_{i,j=1}^3 2 \cos(k_i a)\cos(k_j a)\right]+\ldots \; .\label{eq:quasi}
\end{align}
with the temperature dependent coefficients
\begin{align}
\begin{split}
S_0(T)=&N_S \langle n_k\rangle_k +N_S\frac{z \beta J^2}{U}\left[ \left\langle n_k\frac{n_k(n_l+1)}{n_l-n_k+1} 
\right\rangle_{k,l}-\langle n_k\rangle_k \left\langle \frac{n_k(n_l+1)}{n_l-n_k+1} \right\rangle_{k,l} \right]\quad ,
\end{split}\\
S_1(T)=&N_S \left\langle \frac{2 (n_k+1)n_k}{(n_k-n_l+1)(n_l-n_k+1)}\right\rangle_{k,l}  \quad ,\\
S_2(T)=&N_S \left\langle \frac{n_k(n_l+1)(n_m+1)}{(n_l-n_k+1)(n_m-n_k+1)}  
+\frac{n_k n_l(n_m+1)}{(n_m-n_k+1)(n_m-n_l+1)}\right\rangle_{k,l,m} \quad .
\end{align}
Here $N_S$ denotes the number of lattice sites. Taking an absorption photograph projects the density distribution 
onto a plane. Theoretically, this corresponds to integrating the particle density over the $z$-axis
\begin{align}
n(x,y,t)=\left(\frac{M}{\hbar t}\right)^3\int_{-\infty}^\infty dz \,n\left( \frac{M{\bf r}}{\hbar t}\right)\quad .
\end{align}
Together with a factorization of the Wannier function $w(\mathbf{k})=\prod_{i=1}^3 w(k_i)$ this yields the 
following result for the time-of-flight pictures:
\begin{eqnarray}
&&n(x,y,t)=\frac{M^2}{\hbar^2t^2} \left|w\left(\frac{Mx}{\hbar t}\right)\right|^2\left|w
\left(\frac{My}{\hbar t}\right)\right|^2 \Biggl\{ S_0(T)+2\frac{J}{U} S_1(T) \left[\cos\frac{Max}{\hbar t}
+\cos\frac{May}{\hbar t}+ W(a)\right]+2\frac{J^2}{U^2}S_2(T)\nonumber\\
&&\hspace*{0.5cm}
\times\Biggl[\!\cos\frac{2Max}{\hbar t}\!+\!\cos\frac{2May}{\hbar t}\! +W(2a)\! +4\cos\frac{Max}{\hbar t}
\cos\frac{May}{\hbar t}+4\cos\frac{Max}{\hbar t}W(a)\!+ 4\cos\frac{May}{\hbar t}W(a)\! \Biggr]\Biggr\} 
\!+\ldots\; ,\label{eq:numberint}
\end{eqnarray}
where we have introduced the abbreviation 
\begin{align}
W(l)=\frac{M}{\hbar t}\int_{-\infty}^\infty \left|w\left(\frac{Mz}{\hbar t}\right)\right|^2
\cos\left(\frac{Mlz}{\hbar t} \right)dz \label{eq:wanint} \quad .
\end{align}
Using the Wannier function \eqref{eq:wannier} of the optical lattice within the harmonic approximation 
yields the Fourier transform \cite{Gerbier:PRA05}:
\begin{align}
|w(k_i)|^2=\frac{a}{\pi^{3/2}\sqrt[4]{\tilde{V}_0}}\, \exp\left(-\frac{a^2}{\pi^2 
\sqrt{\tilde{V}_0}}k_i^2 \right)\quad . \label{eq:wannierk}
\end{align}
This allows to calculate \eqref{eq:wanint} explicitly
\begin{align}
W(l)&=e^{-\frac{\pi^2  \sqrt{\Tilde{V}_0}l^2}{4a^2}}  \quad .
\end{align}
In the following we present theoretical density plots of \eqref{eq:numberint} only for the $T\rightarrow0$ limit, i.e.
\begin{align}
S_0(0)&=N_S n \, ,\\
S_1(0)&=2 N_S n(n+1) \, ,\\
S_2(0)&=N_S n(n+1)(2n+1) \, ,
\end{align}
as a procedure to determine the true temperature of bosons in an optical lattice experiment is not yet 
known. Fig.~\ref{fig:tof} compares this result to the experimental findings in 
Refs. \cite{Gerbier:PRL05,Gerbier:PRA05}. There the laser had the wavelength $\lambda=850$ nm  and 
the waist $w=130$ $\mu$m in order to produce a cubic optical lattice, which was superimposed with an external 
harmonic potential of frequency $\omega_m=2\pi\times 15$ Hz. This configuration was filled with $N=2.2\times 10^5$ 
Rubidium-$87$ atoms with an $s$-wave scattering length of $a_{\rm BB}=5.34$ nm, so the recoil energy 
is $E_R=2.10\times10^{-30}$ J. In Appendix \ref{chp:shell} we determine that this experimental situation 
corresponds to the occupation number of $n=2$ bosons per site.
For the high laser intensities in columns (c) and (d) of Fig.~\ref{fig:tof}, where the perturbation parameter $J/U$ 
is sufficiently small, the theoretical pictures match the experimental ones, and there is nearly no difference 
between the first and the second hopping order. In columns (a) and (b), on the other hand, the theory does not fit 
the experiment. In the theoretical picture we can observe the formation of a central and various neighboring peaks, 
which become sharper in the second order, but the neighboring peaks especially are very faint and, furthermore, 
some additional unphysical peaks appear. These problems arise because the perturbation parameter $J/U$ is no longer a 
small quantity in the superfluid regime, so that our approximation is no longer valid. Furthermore, it becomes 
necessary to introduce the  condensate as an order parameter and to take into account its influence on the 
quasi-momentum distribution. These deficiencies can be removed by, for instance, calculating the effective action 
within the hopping expansion \cite{Bradlyn:XXX08}.
\section{Visibility}\label{chp:visibility}
\begin{figure}[t]
{
\psfrag{v}{$\mathcal{V}$}
\psfrag{uj}{$\frac{U}{zJ}$}
\includegraphics{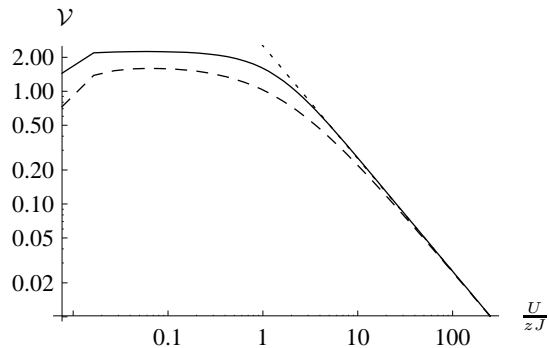}
}
\caption{Comparison of the visibility in first order (dashed) and second order (solid) using Eq.~\eqref{eq:visexp} 
to the first-order expanded result (dotted) given in Eq.~\eqref{eq:vispower} for $T=0$ using the harmonic approximation 
for the Wannier functions.}\label{fig:visibility}
\end{figure}

Following the approach of Refs.~\cite{Gerbier:PRL05,Gerbier:PRA05}, the contrast of peaks in the time-of-flight 
pictures is quantified by the visibility, which is defined as
\begin{align}
\mathcal{V}=\frac{n_{\rm max}-n_{\rm min}}{n_{\rm max}+n_{\rm min}} \quad . \label{eq:visdef}
\end{align}
Here one considers the maximum $n_{\rm max}$ at the first side peak of the integrated density \eqref{eq:numberint}, 
while the minimum $n_{\rm min}$ is taken with the same distance to the central peak.\\
We start with observing that, using the harmonic approximation, the Wannier envelope \eqref{eq:wannierk} cancels 
out in \eqref{eq:visdef}, so that the result only depends on the quasi-momentum distribution \eqref{eq:quasi}. We 
find both the maximum and the minimum by evaluating \eqref{eq:quasi} at the points $x=2\pi\hbar t/Ma,y=0$ 
and $x=y=\sqrt{2}\pi\hbar t/Ma$, respectively. With this both the numerator and denominator of the 
visibility \eqref{eq:visdef} yield in second order of the ratio $J/U$:
\begin{align}
\mathcal{V}=\frac{2(1-\cos\sqrt{2}\pi) S_1(T)\frac{J}{U} +8\sin^2\sqrt{2}\pi  S_2(T)\frac{J^2}{U^2}
+\ldots}{S_0(T)+2(1+\cos\sqrt{2}\pi) S_1(T)\frac{J}{U} +4(1+\cos\sqrt{2}\pi)  S_2(T)\frac{J^2}{U^2}+\ldots} 
\; .\label{eq:visexp}
\end{align}
Expanding this fraction in powers of $zJ/U$ yields
\begin{align}
\mathcal{V}&=v_1(T)\frac{zJ}{U}+v_2(T)\left(\frac{zJ}{U}\right)^{2}+\ldots \label{eq:vispower}\\
\intertext{with the prefactors}
v_1(T)&=\frac{1-\cos(\sqrt{2}\pi)}{3}  \frac{S_1(T)}{S_0(T)} \; , \label{eq:prefactor}\\
v_2(T)&=\frac{\sin^2\sqrt{2}\pi}{9}\left[2\frac{S_2(T)}{S_0(T)}-\left(\frac{S_1(T)}{S_0(T)}\right)^2 \right] ,
\end{align}
which reduce at $T=0$ to
\begin{align}
v_1(0)&=\frac{1-\cos(\sqrt{2}\pi)}{3}(n+1) >0\; ,\label{eq:v1}\\
v_2(0)&=\frac{-2\sin^2\sqrt{2}\pi}{9}(n+1)<0\; .
\end{align}
With this we have generalized Eq.~(5) from Ref.~\cite{Gerbier:PRL05} to finite temperatures with a corrected 
prefactor. Before we discuss the impact of temperature on the visibility, we compare in the $T\rightarrow0$ limit 
Eq.~\eqref{eq:vispower} with direct calculations from Eq.~\eqref{eq:numberint} in Fig.~\ref{fig:visibility}. 
In the double logarithmic plot we use again the occupation number of $n=2$ bosons per site.
All curves show a linear behavior in the Mott phase, while in the superfluid regime our direct 
calculations produce the unphysical result of the visibility being larger than $1$. This is, again, due to fact 
that the above approximations are not appropriate in this parameter region. In the Mott region, on the other hand, 
all curves coincide and the linear dependence in $zJ/U$ is supported by experimental 
data  \cite{Gerbier:PRL05,Gerbier:PRA05}. However, we note that the theoretical prefactor \eqref{eq:v1} 
turns out to be too small by a factor of about $2$.
This discrepancy may arise from the approximation that our calculation is carried out for the homogeneous case, 
so the influence of the shell structure on the visibility is neglected. However, taking this effect into account 
should decrease the theoretical result even further, as the correlation function depends in a nonlinear way on the 
occupation number. In addition, we have also neglected the existence of superfluid regions between the Mott shells, 
but these are small and highly depleted deep in the Mott regime, so this effect should not play a decisive role. \\
Finally, we used the harmonic approximation for the Fourier transform of the Wannier functions so they cancelled 
in \eqref{eq:visexp}. As we show in Appendix \ref{chp:para} that the harmonic approximation can lead to large 
deviations, we investigate now its influence on the visibility.
\begin{figure}[t]
{
\psfrag{V}{$\mathcal{V}$}
\psfrag{UJ}{$\frac{U}{zJ}$}
\includegraphics[width=7cm]{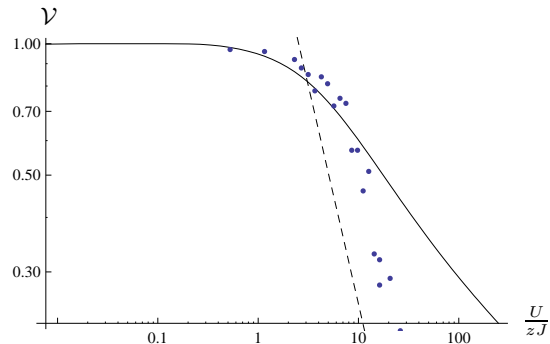}
}
\caption{Comparison of the visibility in the harmonic approximation (dashed) and with numerical determined Wannier 
function in first order (solid) for $T=0$ to the experimental data (dots).}\label{fig:visvergleich}
\end{figure}
We read off from Fig.~\ref{fig:visvergleich}, that the curve with the numerical Wannier envelope in the Mott 
regime no longer has a linear dependence on $J/U$ and lies above the experimental data. This discrepancy between 
our theoretical prediction for the visibility and the experimental findings can be explained as a temperature effect.\\
To this end we investigate at first the expanded result with the harmonic approximation \eqref{eq:vispower}, where 
it is sufficient to study the prefactor \eqref{eq:prefactor} for different chemical potentials, which corresponds to 
different occupation numbers. 
\begin{figure}[t]
{
\psfrag{kTU}{$\frac{k_BT}{U}$}
\psfrag{v}{$v_1$}
\includegraphics{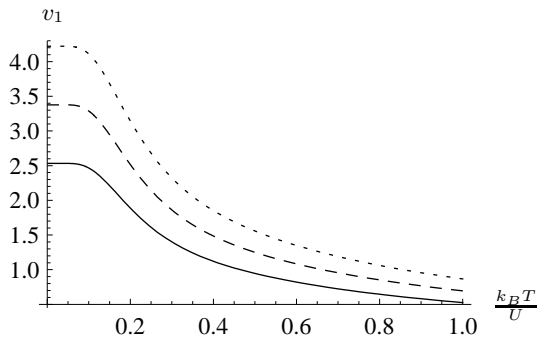}
\caption{Prefactor given by Eq.~\eqref{eq:prefactor} for $\mu/U=1.5$ (solid), $\mu/U=2.5$ (dashed), and $\mu/U=3.5$ 
(dotted).}\label{fig:vistemp}
}
\end{figure}
We see in Fig.~\ref{fig:vistemp}, that the thermal fluctuations slowly destroy the correlation and thus decrease 
the visibility. This qualitative finding within the harmonic approximation remains valid once the Wannier functions 
are calculated numerically.\\
Furthermore, we observe in Fig.~\ref{fig:visvergleich} an increasing discrepancy between our theoretical prediction 
and the experimental findings for larger lattice depths. Note that a detailed analysis shows that this tendency is 
also present in the experimental data \cite{Gerbier:PRA05}. This discrepancy suggests an increasing temperature of 
the bosons in the lattice. This would support the assumption that the ramping of the optical lattice leads to an 
adiabatic heating of the Bose gas \cite{Capogrosso:PRB07,Ho:PRL:07}. In order to take into account this adiabatic 
heating one should determine the relation between temperature~$T$ and the lattice depth $V_0$ from the condition 
that the entropy remains constant throughout the ramping process. 
This task is certainly a nontrivial one as this adiabatic calibration curve $T=T(V_0)$ should be determined across 
the quantum phase transition. We expect that such a calibration curve could be deduced in a consistent way within 
the above mentioned effective action approach \cite{Bradlyn:XXX08}. Once the inhomogeneity of the trap is included, 
it would be possible to get a visibility curve for constant entropy, which could then be used to determine the 
entropy present in the experiment and, finally, to determine the temperature for certain system parameters. One 
could then use remaining deviations from the theory to check if any non-adiabatic heating or non-equilibrium 
final states occur.
\section{Conclusion and Outlook}
In this paper we calculated and analyzed the time-of-flight pictures and the resulting visibility for finite 
temperatures in a perturbative scheme for strong interactions going beyond the commonly used harmonic approximation. 
However, we did not achieve a full understanding of the experimental data, as the experimental situation is quite 
more complex than initially presumed. Comparing our theoretical calculations with experimental data suggested that 
the adiabatic heating process during the ramping of the optical lattice has to be taken into account. Thus, it 
becomes necessary to find a calibration curve, which allows to determine the temperature~$T$ as a function of the 
potential depth~$V_0$ for a given entropy.
\section*{Acknowledgement}
We cordially thank Henrik Enoksen, Fabrice Gerbier, Robert Graham, Konstantin Krutitsky,  Flavio Nogueira, 
Matthias Ohliger, and Ednilson Santos for stimulating discussions. Furthermore, we acknowledge financial support 
from the German Research Foundation (DFG) within the SFB/TR 12 {\it Symmetries and Universality in Mesoscopic Systems}.
\appendix
\section{Parameters of the Bose-Hubbard Model} \label{chp:para}
\begin{figure}[t]
\includegraphics{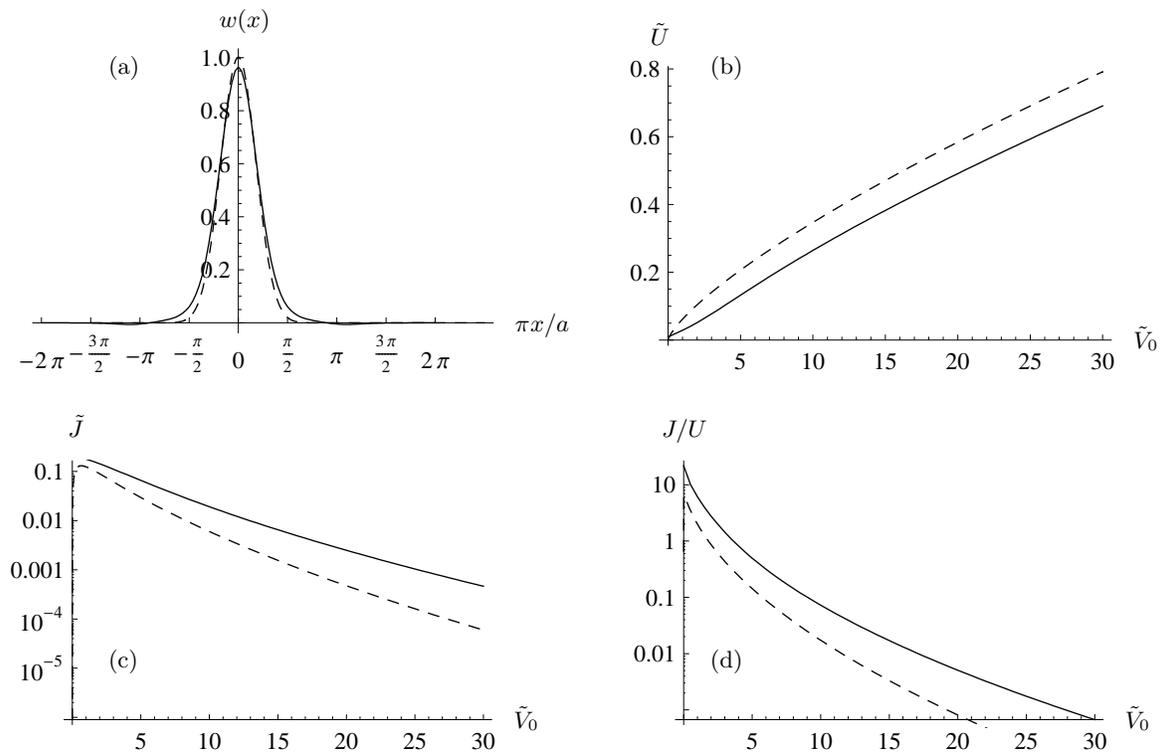}
\caption{Comparison of the harmonic approximation (dashed) and the numerical results (plain) for the Wannier 
function (a), the on-site energy $U$ (b), the hopping energy $J$ (c) and the ratio of the last two (d) using the 
experimental parameter of the Bloch group.} \label{fig:comp}
\end{figure}

We start from the second-quantized Hamiltonian \cite{Jaksch:PRL98}
\begin{align}
\hat{H}=&\int d^3 x \,\hat{\psi}^\dag(\mathbf{x})\left[-\frac{\hbar^2}{2M}\nabla^2+V_{\rm ext}(\mathbf{x})-\mu'  
\right]\hat{\psi}(\mathbf{x}) + \hlf \int d^3 x_1\int d^3 x_2 \, \hat{\psi}^\dag(\mathbf{x}_1) 
\hat{\psi}^\dag(\mathbf{x}_2)\, V_{\rm int}(\mathbf{x}_1,\mathbf{x}_2)\, \hat{\psi}(\mathbf{x}_1) 
\hat{\psi}(\mathbf{x}_2)\; ,  \label{eq:allgham}
\end{align}
where $\hat{\psi}(\mathbf{x})$, $\hat{\psi}^\dag(\mathbf{x})$ are the usual bosonic field operators. The external 
potential is given by the optical lattice $V_{\rm ext}(\mathbf{x})=\sum_{j=1}^3 V_0 \sin^2\left(\pi x_j/a\right) $, 
while we assume that the interaction is of the contact type $V_{\rm int} (\mathbf{x}_1,\mathbf{x}_2) 
=g \;\delta(\textbf{x}_1-\textbf{x}_2)$ with the strength $g= 4 \pi a_{\rm BB} \hbar^2/M$, where $a_{\rm BB}$ 
is the $s$-wave scattering length.\\
As the external potential $V_{\rm ext}(\mathbf{x})$ is periodic, we can expand the field operators in Wannier 
states $w(\mathbf{x}-\mathbf{x}_i)$, which form a complete and orthonormal basis of functions localized at the 
respective sites $i$. As we are at very low temperatures, we restrict ourselves only to the lowest energy band. 
This is implemented by using the decomposition $\hat{\psi}(\mathbf{x})=\sum_i \hat{a }_i w(\mathbf{x}- \mathbf{x}_i)$, 
which yields \eqref{eq:bhm} with the matrix elements:
\begin{align} 
U(i)&=\frac{4 \pi a_{\rm BB} \hbar^2}{M} \int d^3x \, |w(\textbf{x}- \textbf{x}_i)|^4 \label{eq:onsite}\; , \\
J(i,j) &=-\int d^3x \, w^\ast(\textbf{x}- \textbf{x}_i)\, \left[ -\frac{\hbar^2}{2M}\nabla^2+V_{\rm ext} 
(\textbf{x}) \right] \, w(\textbf{x}- \textbf{x}_j) \label{eq:hopp}\quad ,
\end{align}
where the coefficients $U=U(i)$ and $J=J(i,j)$ with $i$ and $j$ being next neighbors turn out to be independent 
of the sites, and the chemical potential is given by $\mu=\mu'+J(i,i)$. Note that due to the orthonormality of the 
Wannier functions $J$ is independent of the dimension while $U$ is not, as the Wannier function factorizes in the 
respective spatial dimensions: $w(\mathbf{x})=\prod_{i=1}^3 w(x_i)$.\\ 
If we use now the harmonic approximation \cite{Zwerger:JoO03,Albus:PRA03} and assume that the ground-state 
wave-function is given by the Gaussian
\begin{align}
w(x_i) =\sqrt[8]{\Tilde{V}_0} \sqrt[4]{\frac{\pi}{a^2}} \exp\left[-\frac{\pi^2}{2}\sqrt{\Tilde{V}_0}
\left(\frac{x_i}{a}\right)^2\right]  \label{eq:wannier} \quad ,
\end{align}
these integrals are directly solved, leading to
\begin{align}
\Tilde{J}&=\left(\frac{\pi^2}{4} - 1\right)\Tilde{V}_0\; e^{-\pi^2 \sqrt{\Tilde{V}_0} /4 }\; , \\
\tilde{U}&=\sqrt{8 \pi} \frac{a_{\rm BB}}{a}\Tilde{V}_0^{3/4} \quad .
\end{align}
Alternatively, we can determine the Wannier functions numerically \cite{Krutitsky:PC} and use this result to 
calculate the on-site interaction strength $U$ via \eqref{eq:onsite}. In contrast, the tunneling parameter $J$ 
should not be determined from \eqref{eq:hopp} as numerical differentiation leads to low precision. Instead we use 
the fact that $J$ does not depend on the dimension and the definition of the Wannier functions 
$w(x- x_i)=N_S^{-1/2}\sum_{k} e^{-i  k x_i}\phi_{k}(x)$, with the single particle Schr\"odinger wave function 
of the lowest band $\phi_{k}(x)$ and the corresponding eigenenergy $E(k)$. With this \eqref{eq:hopp} 
reduces to \cite{Blakie:JoO04}
\begin{align}
J&=\frac{1}{N_S} \sum_{k} e^{ik(x_i- x_j)} E(k)\quad .
\end{align}
When we compare the analytical approaches with the numerical one in Fig.~\ref{fig:comp}(a), we find, that the 
harmonic approximation to the Wannier function nearly fits the numerical counterpart but for one special feature: 
a Gaussian is always positive while the numerical curve shows oscillations around zero. Note that it is, in fact, 
indispensable that the Wannier functions have also negative values in order to guarantee their orthogonality. This 
discrepancy leads to a significant deviation of the qualitative behavior in both 
Bose-Hubbard parameters over the whole 
energy scale. In the on-site energy $U$ in Fig.~\ref{fig:comp}(b) the harmonic approximation yields values, which 
are too large, but the relative error is at least slowly decreasing for higher laser intensities, while the relative 
error of the hopping parameter $J$ in Fig.~\ref{fig:comp}(c) is even growing. The ratio of these two given in 
Fig.~\ref{fig:comp}(d) is the central calibration curve to connect the experimental data to the theory. Because 
of the higher precision, we use this numerical calibration curve throughout our paper.
\section{Shell Structure}\label{chp:shell}
\begin{figure}[t]
\psfrag{i}{$i$}
\psfrag{n}{$n$}
\includegraphics[width=7cm]{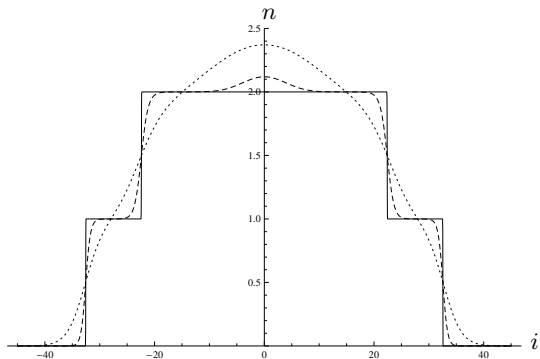}
\caption{Shell structure slice of the Mott-phase for $T=0$ (solid), $k_BT/U=0.05$ (dashed), and 
$k_BT/U=0.2$ (dotted).}\label{fig:denstrap}
\end{figure}

For our consideration it is important to know how many particles are at each site depending on the total particle 
number. In theory this number for $T=0$ is given by the next integer to the ratio $\mu/U$.\\
In the homogeneous case, however, there is no natural restriction to the number of sites. Therefore, for any number 
per site one can find a number of sites, which gives the desired total particle number. In the experiments, however, 
we have an additional overall harmonic trapping potential with a frequency $\omega$ present. Additionally one has to 
take into account that the harmonic trap is enhanced by the shape of the laser beam. As the laser beam has a Gaussian 
intensity distribution, the lattice potential is deeper in the center of the laser beam than at the edge. The 
influence of this laser inhomogeneity can be described by an effective frequency \cite{Gerbier:ARX07}:
\begin{align}
\omega \approx \sqrt{\omega_m^2+\frac{8 V_0 -4 E_R \sqrt{V_0/E_R}}{m w^2}} \label{frm:oeff}\quad .
\end{align}
which limits the number of occupied sites. Additionally, it changes the local chemical potential and leads to 
concentric shells of Mott insulators, which have different numbers of particles per site. This is known as the 
wedding cake structure, which has been confirmed by comparing theoretical predictions \cite{Batrouni:PRL02} with 
experimental results \cite{Folling:PRL06}. A non-vanishing 
temperature blurs the boundaries between this shells as depicted 
in Fig.~\ref{fig:denstrap}.
\begin{figure}[t]
{
\psfrag{N}{$N$}
\psfrag{mu}{$\mu/U$}
\includegraphics[width=7cm]{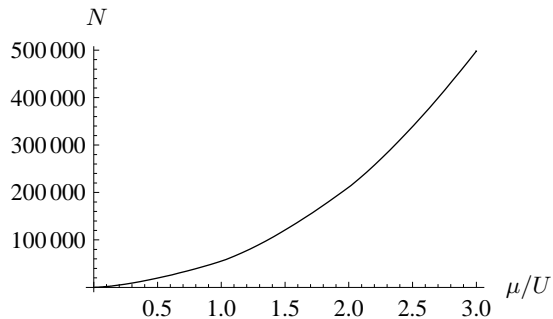}
}
\caption{Calibration curve for the chemical potential in the experiment of the Bloch group.}\label{fig:chempot}
\end{figure}
As it is possible to measure this density variations one could use the shape of the boundaries to 
determine the temperature. But in this picture we neglected the superfluid shells, which are between the 
Mott shells. They are present even at $T=0$ and produce a similar blurring.\\
If we neglect the discrete nature of the sites, determine the radii of the layers and subsequently add their 
volumes, we obtain for the equation of state \cite{Ho:PRL:07}
\begin{align}
N=&\sum_{0\leq c<\mu/U} \frac{4\pi}{3}\sqrt{\frac{\mu-c U}{\hlf m \omega^2a^2}}^3 \quad .
\end{align}
This yields the calibration curve of Fig.~\ref{fig:chempot}, which has only negligible deviations from the exact 
sum over all lattice sites. From the calibration curve we can determine from a given particle number $N$ the chemical 
potential $\mu$ for the experiment and from this the occupation number in the central Mott shell. As the experiments 
of the Bloch group are done with a total particle numbers of about $N=2.2\times10^5$, the chemical potential is slightly 
smaller than $2 U$. Hence, the maximal occupation number is $2$, which is consistent with the claims 
of Ref.~\cite{Gerbier:PRA05}.

\begin{thebibliography}{99}
%
\bibitem{Lewenstein:AdP07}
M. Lewenstein, A. Sanpera, V. Ahufinger, B. Damski, A. Sen(De), and U. Sen,
Adv. Phys. {\bf 56}, 243 (2007).
%
\bibitem{Bloch:RMP08}
I. Bloch, J. Dalibard, and W. Zwerger,
Rev. Mod. Phys. {\bf 80}, 885 (2008).
%
\bibitem{Fisher:PRB89}
M. P. A. Fisher, P. B. Weichman, G. Grinstein, and D. S. Fisher,
Phys. Rev. B {\bf 40}, 546 (1989). 
%
\bibitem{Jaksch:PRL98}
D. Jaksch,  C. Bruder, J. I. Cirac, C. W. Gardiner, and P. Zoller,
Phys. Rev. Lett. {\bf 81}, 3108 (1998).
%
\bibitem{Greiner:Nat02}
M. Greiner, O. Mandel, T. Esslinger, T. W. H{\"a}nsch, and I. Bloch,
Nature {\bf 415}, 39 (2002).
%
\bibitem{Bloch:NaP05}
I. Bloch,
Nature Phys. {\bf 1}, 23 (2005).
%
\bibitem{Zwerger:ASS04}
W. Zwerger,
Adv. Sol. State Phys. {\bf 44}, 277 (2004).
%
\bibitem{Zoller:AP(NY)05}
D. Jaksch and P. Zoller,
Ann. Phys. (New York) {\bf 315}, 52 (2005). 
%
\bibitem{Stoeferle:PRL04}
T. St\"{o}ferle, H. Moritz, C. Schori, M. K\"{o}hl, and T. Esslinger,
Phys. Rev. Lett. {\bf 92}, 130403 (2004).
%
\bibitem{Folling:PRL06}
S. F\"olling, A. Widera, T. M\"uller, F. Gerbier, and I. Bloch,
Phys. Rev. Lett. {\bf 97}, 060403 (2006).
%
\bibitem{Gerbier:PRA05}
F. Gerbier, A. Widera, S. F{\"o}lling, O. Mandel, T. Gericke, and I Bloch,
Phys. Rev. A  {\bf 72}, 053606 (2005).
%
\bibitem{Morsch:RmP06}
O. Morsch and M. Oberthaler,
Rev. Mod. Phys {\bf 78}, 179 (2006). 
%
\bibitem{Greiner:Nat02.A}
M. Greiner, O. Mandel, T.W. H{\"a}nsch, and I. Bloch,
Nature, {\bf 419}, 51 (2002).
%
\bibitem{Kollath:PRL07}
C. Kollath, A. M. L\"{a}uchli, and E. Altman,
Phys. Rev. Lett. {\bf 98}, 180601 (2007).
%
\bibitem{Schutzhold:ARX08}
U. R. Fischer and R. Sch\"utzhold,
eprint: {\tt arXiv:0807.3627}.
%
\bibitem{Rom:Nat06}
T. Rom, Th. Best, D. van Oosten, U. Schneider, S. F{\"o}lling, B. Paredes, and I. Bloch,
Nature {\bf 444}, 733 (2006).
%
\bibitem{Esslinger:PRL06}
K. G\"unter, T. St\"oferle, H. Moritz, M K\"ohl, and T. Esslinger,
Phys. Rev. Lett. {\bf 96}, 180402 (2006).
%
\bibitem{Ospelkaus:PRL06}
S. Ospelkaus, C. Ospelkaus, O. Wille, M. Succo, P. Ernst, K. Sengstock, and K. Bongs,
Phys. Rev. Lett. {\bf 96}, 180403 (2006).
%
\bibitem{Refael:PRB08}
G. Refael and E. Demler,
Phys. Rev. B {\bf 77}, 144511 (2008).
%
\bibitem{Ospelkaus:PRL06.B}
C. Ospelkaus, S. Ospelkaus, L. Humbert, P. Ernst, K. Sengstock, and K. Bongs,
Phys. Rev. Lett. {\bf 97}, 120402 (2006).
%
\bibitem{Damski:PRL03}
B. Damski, L. Santos, E. Tiemann,  M. Lewenstein, S. Kotochigova, P.  Julienne,  and P. Zoller,
Phys. Rev. Lett. {\bf 90}, 110401 (2003).
%
\bibitem{Krutitsky:NJP06}
K. V. Krutitsky, A. Pelster, and R. Graham,
New J. Phys. {\bf 8}, 187 (2006).
%
\bibitem{Krutitsky:08}
K. V. Krutitsky, M. Thorwart, R. Egger, and R. Graham,
Phys. Rev. A {\bf 77}, 053609 (2008).  
%
\bibitem{Hofstetter:08}
U. Bissbort and W. Hofstetter,
eprint: {\tt arXiv:0804.0007}.
%
\bibitem{Garcia:TRS03}
J. J. Garcia-Ripoll and J. I. Cirac,
Phil. Trans. R. Soc. {\bf 361}, 1537 (2003). 
%
\bibitem{Kay:NJP06}
A. Kay and K. Pachos,
New J. Phys. {\bf 6}, 126 (2004). 
%
\bibitem{Treutlein:FPh06}
P. Treutlein, T. Steinmetz, Y. Colombe, B. Lev, P. Hommelhoff, J. Reichel, 
M. Greiner, O. Mandel, A. Widera, T. Rom, I. Bloch, T. W. H\"ansch,
Fortschr. Phys. {\bf 54}, 702 (2006). 
%
\bibitem{Oosten:PRA01}
D. van Oosten,  P. van der Straten, and H. T. C. Stoof,
Phys. Rev. A {\bf 63}, 053601 (2001).
%
\bibitem{Freericks:PRB96}
J. K. Freericks and H. Monien,
Phys. Rev. B {\bf 53}, 2691 (1996).
%
\bibitem{Damski:PRA06}
B. Damski and J. Zakrzewski,
Phys. Rev. A {\bf 74}, 043609 (2006).
%
\bibitem{Batrouni:PRL95}
G. G. Batrouni, R. T. Scalettar, and G. T. Zimanyi,
Phys. Rev. Lett. {\bf 65}, 1765 (1990).
%
\bibitem{Capogrosso:PRB07}
B. Capogrosso-Sansone, N. V. Prokof'ev, and B. V. Svistunov,
Phys. Rev. B {\bf 75}, 134302 (2007).
%
\bibitem{Capogrosso:PRA08}
B. Capogrosso-Sansone, S. G. S\"oyler, N. V. Prokof'ev, and B. V. Svistunov,
Phys. Rev. A {\bf 77}, 015602 (2008).
%
\bibitem{Santos:ARX08}
E. F. A. dos Santos and A. Pelster,
eprint: {\tt arXiv:0806.2812}.
%
\bibitem{Buonsante:PRA04}
P. Buonsante and A. Vezzani,
Phys. Rev. A {\bf 70}, 033608 (2004).
%
\bibitem{Gerbier:PRL07}
F. Gerbier,
Phys. Rev. Lett. {\bf 99}, 120405 (2007).
%
\bibitem{Cramer:ARX07}
M. Cramer, S. Ospelkaus, C. Ospelkaus, K. Bongs, K. Sengstock, and J. Eisert,
Phys. Rev. Lett. {\bf 100}, 140409 (2008). 
%
\bibitem{Ziegler:94}
K. Ziegler,
Physica A {\bf 208}, 177 (1994). 
%
\bibitem{Ziegler:02}
K. Ziegler,
J. Low Temp. Phys. {\bf 126}, 1431 (2002).
%
\bibitem{Ziegler:03}
K. Ziegler,
Las. Phys. {\bf 13}, 587 (2003).
%
\bibitem{Zwerger:JoO03}
W. Zwerger,
J. Optics B {\bf 5}, S9 (2003).
%
\bibitem{Albus:PRA03}
A. Albus, F. Illuminati, and J. Eisert,
Phys. Rev. A {\bf 68}, 023606 (2003).
%
\bibitem{Karshunikov:PRA02}
V. A. Kashurnikov, N. V. Prokof'ev, and B. V. Svistunov,
Phys. Rev. A {\bf 66}, 031601(R) (2002).
%
\bibitem{Gerbier:PRL05}
F. Gerbier, A. Widera, S. F\"olling, O. Mandel, T. Gericke, and I Bloch,
Phys. Rev. Lett. {\bf 95}, 050404 (2005).
%
\bibitem{Bradlyn:XXX08}
B. Bradlyn, E. F. A. dos Santos, and A. Pelster,
{\tt arXiv:0809.0706}.
%
\bibitem{Ho:PRL:07}
T.-L. Ho and Q. Zhou
Phys. Rev. Lett. {\bf 99}, 120404 (2007).
%
\bibitem{Krutitsky:PC}
K. Krutitsky,
private communication.
%
\bibitem{Blakie:JoO04}
P. B. Blakie and C. W. Clark,
J. Phys. B {\bf 37}, 1391 (2004).
%
\bibitem{Gerbier:ARX07}
F. Gerbier, S. F\"olling, A. Widera, and I. Bloch,
eprint: {\tt cond-mat/0701420}.
%
\bibitem{Batrouni:PRL02}
G. G. Batrouni,  V. Rousseau, R. T. Scalettar,  M. Rigol, A.  Muramatsu, P. J. H. Denteneer, and M. Troyer,
Phys. Rev. Lett. {\bf 89}, 117203 (2002).
%
\end{thebibliography}
\end{document}